# How is a gas sensor poisoned by volatile methylsiloxanes?

Heng Liu[1,†], Bingxin Yang[2,†], Yiming Lu[1,†], Yuan Wang[1], Xue Jia[1], Long Luo[2,*], Hao Li[1,*]

[1] *Advanced Institute for Materials Research (WPI-AIMR), Tohoku University, Sendai 980-8577, Japan*

[2] *Department of Chemistry, University of Utah, Salt Lake City, UT 84112, USA*

[†] *These authors contribute equally.*

Corresponding Author Email:
long.luo@utah.edu (L.L.)
li.hao.b8@tohoku.ac.jp (H.L.)

## Abstract

Volatile methyl siloxanes (VMSs), widely present in consumer and industrial products, have attracted increasing concerns due to their persistence, bioaccumulation behavior, and adverse health effects. Beyond their environmental implications, VMSs also pose operational challenges for sensing technologies because they readily decompose on sensing materials to form silicon-based compounds (*e.g.*, silica and silane) that irreversibly impair sensing performance, a phenomenon commonly known as **"siloxane poisoning"**. Despite its prevalence, the mechanistic basis of this deactivation remains poorly understood. **Herein, we present the first comprehensive theoretical study of siloxane-induced poisoning in catalytic gas sensors.** Guided by our self-developed AI Agent, Digital Sensor Platform (*DigSen*), we first identify siloxane poisoning as a previously overlooked yet high-impact research direction. Using hexamethyldisiloxane (HMDS) as a model compound, we then conducted first-principles calculations to uncover decomposition pathways across noble metal surfaces. Strikingly, a descriptor-based microkinetic volcano model is developed to capture the trade-off between sensing activity and resistance to poisoning, enabling predictive identification of anti-poisoning candidates. These insights not only elucidate the origin of siloxane poisoning but also demonstrate how AI-driven discovery, mechanistic theory, and experiments can be integrated into a closed-loop framework for catalytic sensor design. More broadly, this AI-guided paradigm represents a generalizable strategy for materials digital discovery, offering a transferable methodology that extends well beyond siloxane systems to diverse classes of materials challenges.

**Introduction:**

Siloxanes are organosilicon compounds composed of silicon atoms bonded to at least one carbon and one oxygen atom.[1] Due to their hydrophobicity and excellent biocompatibility, siloxanes are widely used in various applications, including construction (*e.g.*, concrete sealers, brick sealers, and masonry water repellents), electronics, automotive manufacturing, textiles, medical devices, food packaging, cookware, personal care products (*e.g.*, hair gels, shampoos, facial creams), and cosmetics (*e.g.*, breast implants).[2] According to literature reports, global siloxane production has reached 8–10 million tons per year.[3]

Despite the growing number of studies, current knowledge of siloxanes remains highly fragmented across domains such as environmental science, catalysis, and sensor technology. As a result, the interconnections between siloxane degradation, catalytic surface reactions, and device deactivation remain poorly understood. To systematically capture and analyze this dispersed body of knowledge, we developed a bespoke AI Agent, Digital Sensor Platform (*DigSen*: www.digsen.org), based on our comprehensive literature analysis methodology. By constructing a curated corpus of siloxane-related literature and implementing a domain-specific large language model (LLM) pipeline, the *DigSen* systematically extracted, organized, and interpreted information on the use, degradation, and material impact of siloxanes.

Strikingly, through this automated discovery process, the *DigSen* AI Agent revealed catalytic siloxane decomposition and its consequential surface poisoning in gas sensors as a critical yet underexplored problem linking environmental chemistry and device performance. The analysis revealed that volatile methyl siloxanes (VMSs) with low to medium molecular weights (<500 g/mol)[2] are frequently detected in various environmental media, including air, dust, water, soil, sediment, and sludge.[4] Multiple studies have highlighted the potential toxicity of VMSs to living organisms, including adverse effects on reproductive systems, carcinogenicity, and respiratory harm.[4, 5] Consequently, VMSs are now regarded as emerging environmental pollutants and are subject to increasing regulatory scrutiny.[6]

Beyond environmental concerns, *DigSen* AI Agent also revealed that the oxidation of VMSs to $SiO_2$ poses operational challenges in industrial systems. The accumulation of silica particles inevitably impairs equipment and device performance, most notably by deactivating sensors due to siloxane poisoning. The decomposition of hexamethyldisiloxane (HMDS), a commonly studied VMS, results in pore clogging and irreversible deactivation of gas sensors.[7, 8] Several mitigation strategies have been explored to reduce catalyst poisoning. These include doping noble metals with transition metals—*e.g.*, adding Fe to Pt/$\gamma$-$Al_2O_3$ catalysts has shown some ability to improve HMDS tolerance, although it does not fully prevent deactivation.[9] Surface modification techniques have also been proposed. For instance, Chilton et al. demonstrated that activated charcoal and carbon cloth adsorbents can extend the operational life of methane sensors.[10] However, a systematic understanding of how volatile siloxanes react on a catalytic metal surface is still lacking.

Motivated by the above AI-driven insights, we herein investigate the catalytic decomposition of VMSs using HMDS as a model compound, employing an integrated theoretical and experimental approach. Our findings provide valuable insights into the development of siloxane-tolerant catalysts and offer a promising foundation for new catalytic strategies aimed at removing siloxanes from the environment. Moreover, this work provides a foundational framework for the rational development of siloxane-tolerant catalytic gas sensors and illustrates the potential of AI- and theory-guided discovery in uncovering emerging challenges in sensor applications.

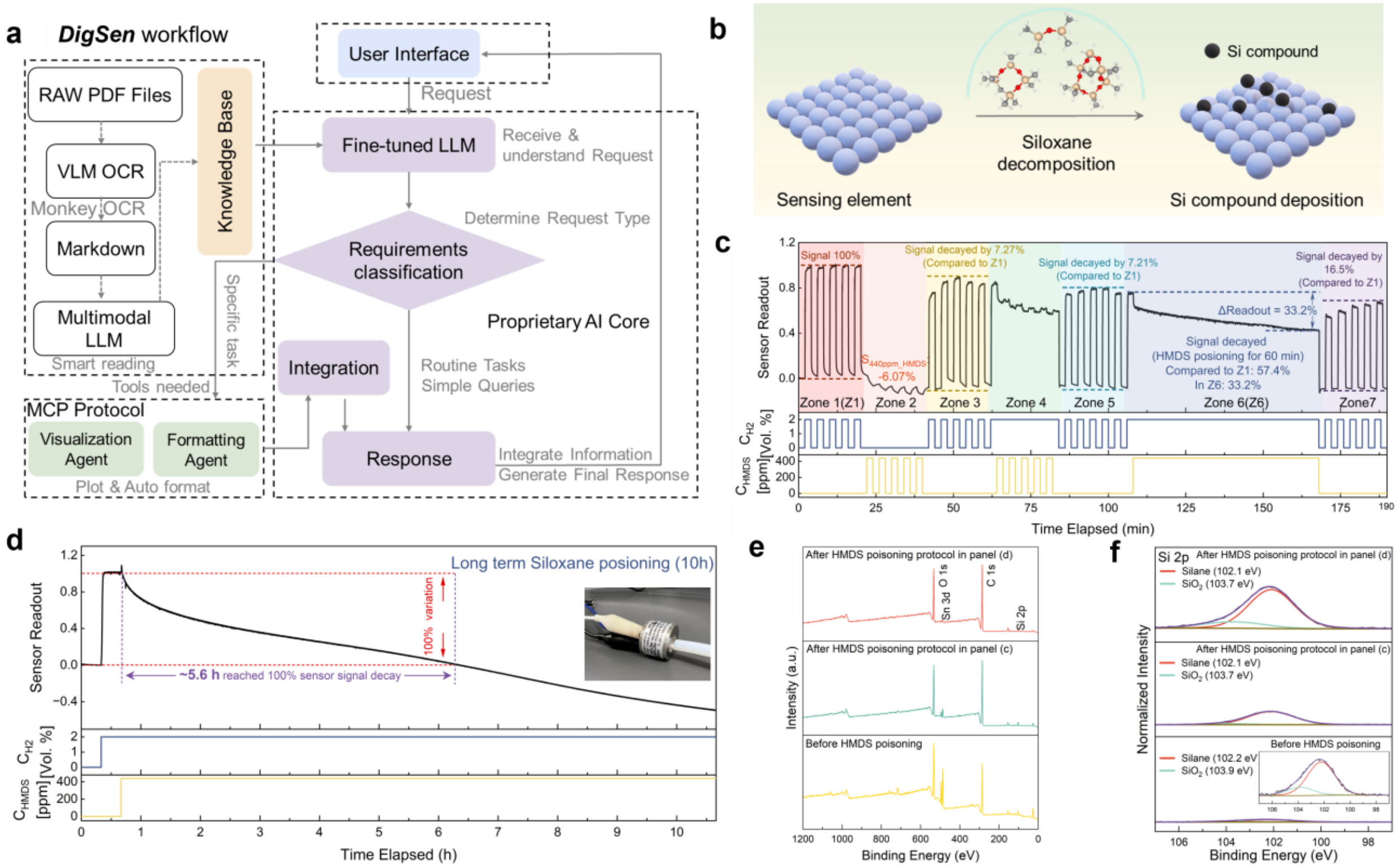

**Figure 1. AI-empowered identification of siloxane poisoning as an important yet underexplored research direction and experimental proof of concept.** a, Workflow of the self-developed AI Agent, *DigSen*, used to uncover siloxane-related research opportunities. b, Schematic diagram of Si-containing species deposition on the sensing element as a result of catalytic siloxane decomposition. c, Response of a commercial catalytic combustible gas sensor when exposed to air containing HMDS, $H_2$, or both. d, Response of a commercial catalytic gas sensor to 2% $H_2$ gas in the presence of 440 ppm HMDS for over 10 hours. e, XPS survey spectra before and after HMDS poisoning. f, High-resolution XPS spectra in the Si 2*p* region of the commercial catalytic sensor before and after HMDS poisoning tests in panels (c) and (d) (The inset in **Figure 1f** shows a magnified view of Si 2*p* spectrum before HMDS poisoning).

## Results and Discussions

To systematically investigate siloxane-induced poisoning of sensing materials, we developed the first domain-specific AI Agent for catalytic gas sensing, *DigSen* (version for public use: www.digsen.org), built on a curated knowledge base of the most comprehensive siloxane-related literature reported to date (**Figure 1a**). By leveraging LLMs and Vision Language Model (VLM), it allows for the precise extraction of not only text but also complex elements like tables and scientific formulas by first identifying their structure, then recognizing their content, and finally mapping their logical relationships within the document. The technological details are provided in **Figure S1,** and the basic techniques for constructing this AI Agent are provided in **Supplementary Information**. *DigSen* is developed as an asset MCP module that can be smoothly integrated into supported clients like Cursor and CherryStudio. This allows it to serve as a more versatile frontend compared to our online version. With the assistance of AI Agents, the automated discovery directly revealed a critical insight: while the environmental accumulation and toxicity of VMSs have been extensively studied, there is a conspicuous lack of mechanistic understanding regarding their catalytic decomposition and the resulting sensor poisoning phenomenon. In this context, the AI Agent played a central role in identifying an underexplored scientific direction and framing siloxane-induced deactivation in catalytic sensors as a distinct research problem. This approach demonstrates the potential of AI tools not only for knowledge integration but also for digesting complex knowledge from literature *via* their strong "thinking power" and guiding theoretical exploration in emerging areas.

To further address the knowledge gap, we conducted a series of proof-of-concept experiments using commercial gas sensors. This general poisoning mechanism is schematically depicted in **Figure 1b**, where silicon-containing species generated from siloxane decomposition deposition (*e.g.*, $SiO_2$ or silane) deposit on the surface of the sensing element and competitively occupy active sensing sites, ultimately leading to significant performance degradation. To experimentally validate it, we exposed a commercial gas sensor to various combinations of the target analyte ($H_2$) and the poisoning gas (*i.e.*, HMDS), as shown in **Figure 1c**. Relevant methods are detailed in **Supplementary Information**. Initially, the sensor was exposed to five cycles of 2% $H_2$ in air (**Zone 1**), producing highly reproducible responses that were averaged and defined as the 100% reference signal. When exposed to 440 ppm HMDS in air (**Zone 2**), the sensor exhibited an average signal change of –6.07%, suggesting that HMDS significantly alters the surface chemistry of the sensing material. After five successive exposures to 440 ppm HMDS, the response to 2% $H_2$ in air decreased by an average of 7.22% (**Zone 3**) compared to the **Zone 1** reference, indicating clear poisoning. Subsequent exposure to a mixture of 2% $H_2$ and 440 ppm HMDS pulses (**Zone 4**) caused a pronounced decay in the sensor signal, with sluggish recovery after each pulse. Following this mixed-gas exposure, the response to 2% $H_2$ (**Zone 5**) further decreased, and the sensor did not return to its original baseline once $H_2$ was removed. In **Zone 6**, extended exposure to the $H_2$/HMDS mixture for 1 hour resulted in an additional 33.2% decline in the sensing signal. Finally, removing HMDS while maintaining $H_2$ exposure led to partial recovery of the sensing signal (**Zone 7**), suggesting that a fraction of silicon-containing poisoning species could be removed from the sensor surface. To evaluate the impact of long-term siloxane poisoning, the sensor was first exposed to 2% $H_2$ for 20 mins and then continuously to a 2% $H_2$ /440 ppm HMDS mixture for 10 hours. The results, shown in **Figure 1d**, demonstrate substantial degradation of $H_2$ sensing performance, with complete signal extinction (from 100% reference to 0%) within ~6 h.

To elucidate the underlying surface chemical changes, X-ray photoelectron spectroscopy (XPS) analysis was performed on the commercial catalytic gas sensor before and after siloxane poisoning tests. The spectra were calibrated by aligning the adventitious carbon peak to 284.8 eV. As shown in **Figure 1e**, survey scans after HMDS poisoning protocols in **Figures 1c** and **1d** revealed the presence of Si on the $SnO_2$ sensing material. To further explore the Si species depositions, our high-resolution Si 2*p* spectra (**Figure 1f**), which was normalized to the Sn 3*d* peak areas, displayed two deconvoluted peaks with increased intensity after HMDS exposure. The peaks at 102.1 and 103.7 eV were assigned to silane and $SiO_2$, respectively.[11, 12] In addition, the O 1*s* (**Figure S4a**) spectra also support the existence of silicon species. The peak at 531.9 eV corresponds to Si–O–Si (bridging oxygen in siloxane), while the higher binding energy component at 533.4 eV is the characteristic of $SiO_2$.[13-16] Moreover, the C 1s (**Figure S4b**) spectra show that binding energies located at 284.8, 286.1, and 288.6 eV are associated with C-C, C-O, and C=O, respectively,[17] indicating the presence of methyl-terminated siloxane on the sensing material surface. Furthermore, we quantified the ratio of Si to Sn in **Table S1**, the latter one being from the sensing element of the commercial sensor. The ratios were determined to be 1.05, 3.71, and 10.6 for the sensing material before HMDS poisoning, after HMDS poisoning protocol in panel d, and after HMDS poisoning protocol in panel e, respectively. The increasing ratio suggests an increasing Si species deposition on the sensing element. **These results support that the loss in sensor performance originates from HMDS decomposition to form silicon species including silica and silane on the sensing surface.**

To build the structure-performance relationship of the HMDS decomposition, density functional theory (DFT) calculations (detailed computational methods can be referred in **Supplementary Information**) were performed to study the decomposition mechanism of HMDS, beginning with an investigation of its interaction with catalytic metal surfaces. To this end, atomic models of HMDS were constructed in both its isolated gas-phase configuration and its adsorbed states on representative catalytic surfaces, including Pt(111), Pd(111), Au(111), Ir(111), and Ag(111). The reason for choosing (111) surfaces for analysis is because it is known to be the most stable surface for these face-centered cubic transition metals due to its close-packed nature. An example of HMDS adsorbed on Pt(111) is depicted in **Figure 2a**. The inset shows the charge density difference between Pt(111) and adsorbed HMDS, which indicates a strong electronic interaction upon adsorption. The calculated charge density differences between other materials (Pd/Au/Ir/Ag) and HMDS are presented in **Figure S5**, revealing the same conclusion. These interfacial interactions are essential for triggering the decomposition of HMDS, a phenomenon that underlies sensor poisoning and silica deposition observed in industrial systems.

To quantify this interfacial bond perturbation, we performed crystal orbital Hamilton population (COHP) analyses[18] for three representative bonds in HMDS, that is Si–C, C–H, and Si–O, under both gas-phase and adsorbed conditions. As illustrated in **Figure 2b**, gas-phase HMDS displays stronger bonding character with integrated COHP (-ICOHP) values of 2.45 eV for Si–C, 2.77 eV for C–H, and 3.46 eV for Si–O. Among them, the Si–C bonding is the weakest, implying its potential susceptibility to activation. When HMDS is adsorbed on Pt(111) (**Figure 2c**), all three bonds experience weakening due to electronic interactions with the metal surface. The Si–C bond undergoes the most substantial reduction in ICOHP, dropping to 1.94 eV, suggesting it as the most activated and hence likely to dissociate first. This finding aligns well with prior experiments identifying Si–C cleavage as the initial step in HMDS decomposition on noble metal catalysts.[7, 19] The COHP evolution on Pd(111), Au(111), Ir(111), and Ag(111) present similar trends, as illustrated in **Figures S6-S9**.

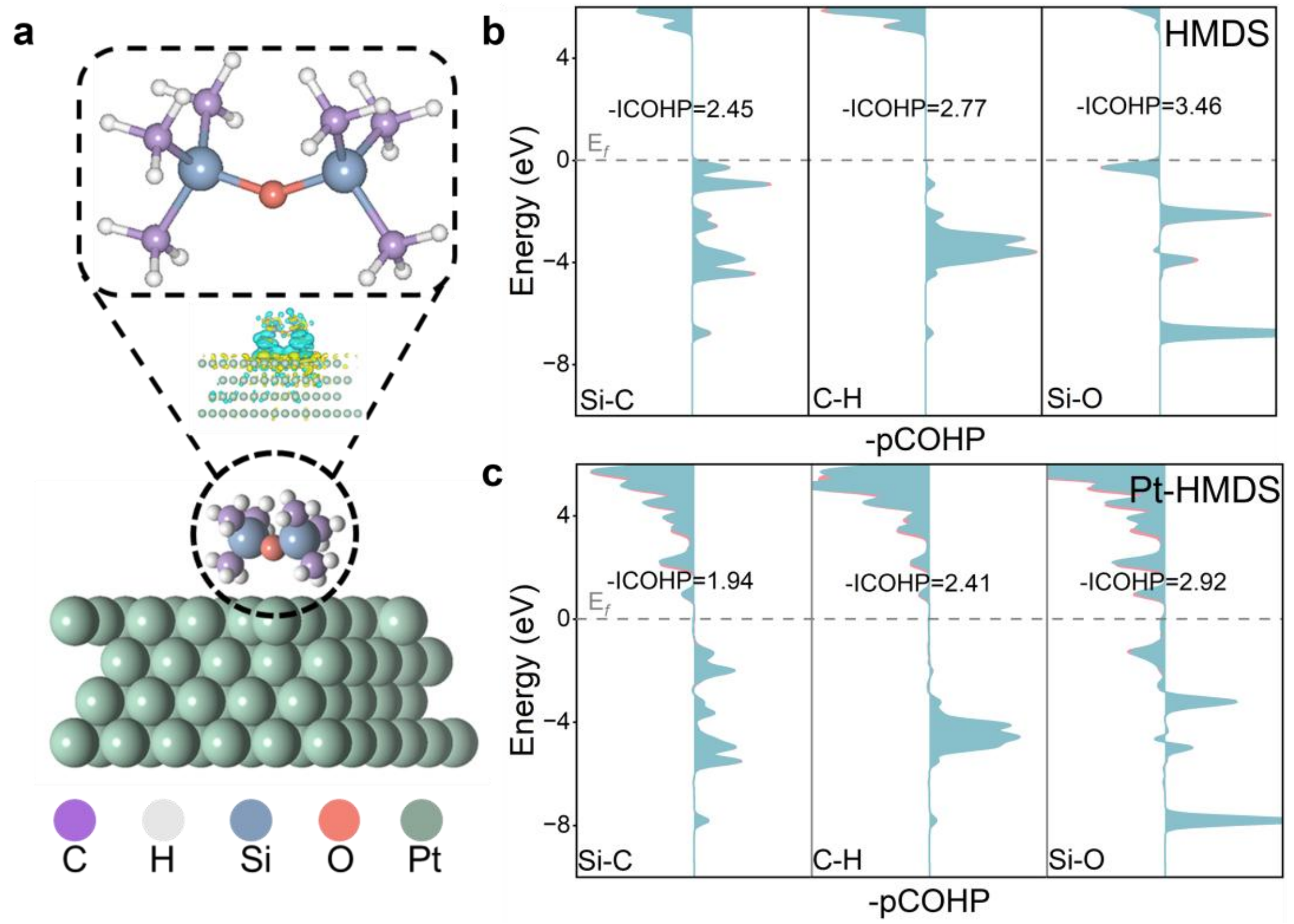

**Figure 2**. **COHP analysis of HMDS in gas phase and on Pt surface. a,** Atomic structures of HMDS in gas phase and adsorbed on Pt(111) surface. **b,** Projected crystal orbital Hamilton population (-pCOHP) and integrated COHP (-ICOHP) for Si–C, C–H, and Si–O bonds in the gas-phase HMDS molecule. **c,** Corresponding -pCOHP and -ICOHP plots for HMDS adsorbed on Pt(111), revealing bond weakening induced by surface

interaction. Color codes: Si (blue), H (white), C (purple), O (red), Pt (green). Inset shows the charge density difference between adsorbed HMDS and Pt(111), in which cyan and yellow regions signify charge depletion and accumulation. Isosurfaces refer to 1.62´$10^{-4}$ electrons/$bohr^3$.

Building upon the bond strength hierarchy established in **Figure 2**, we proceeded to evaluate the thermodynamic feasibility of HMDS decomposition on different metal surfaces by calculating the reaction energies (ΔE) associated with key bond dissociation events. To note, we focused exclusively on the scission of Si–C and Si–O bonds—the two primary linkages that define the siloxane backbone. The comprehensive decomposition pathways for Pt(111), Pd(111), Au(111), Ir(111), and Ag(111) are shown in **Figure 3**. Obviously, each pathway initiates with the cleavage of a Si–C bond, with the ΔE of -1.72 eV (Pt), -0.82 eV (Pd), -0.33 eV (Au), -1.24 eV (Ir), and 0.34 eV (Ag) much lower than those of Si-O, which is in consistent with the above COHP analysis. All subsequent steps follow the same trend until the 5$^{th}$ step (marked with diamond frame) for Ag and 6$^{th}$ step for Pd, Au, and Ir (in square frame). Specifically, when decomposed to *Si-O-Si-$CH_3$ (where * denotes the surface site), Si-O on methyl-side will be broken on Pd(111), Au(111), and Ir(111), forming *SiO and *$SiCH_3$, followed by a final Si-C breaking. As for the case in Ag, after forming the symmetric *$H_3C$-Si-O-Si-$CH_3$, cleavage will happen on Si-O bond, followed by two consecutive Si-C bond breaking steps. Detailed energy information for each step on various materials are provided in **Figures S10-S14**.

To further explore the mechanistic divergence observed in the 5$^{th}$ and 6$^{th}$ decomposition stages, projected density of states (PDOS) and Bader charge analyses were implemented on *$H_3C$-Si-O-Si-$CH_3$ for all the metal surfaces, and *Si-O-Si-$CH_3$ for Pt, Pd, Au, and Ir. The focus was placed on the electronic interaction between the Si atom bonded to the methyl group and the bridging O atom, which plays a central role in determining the preferred bond-breaking pathway. For *$H_3C$-Si-O-Si-$CH_3$ (the 5$^{th}$ step), as presented **Figure 4**, surrounded by black dash frame, the two surface-anchored Si atoms exhibit a relatively lower net charge (-1.62 e) on Ag(111), indicating weaker electron back-donation from the metal surface. In contrast, stronger charge accumulation on Si atoms (ranging from –2.01 to –2.49 e) is observed on Pt, Pd, and Ir surfaces, reflecting more extensive metal–adsorbate charge transfer. This variation in charge distribution reflects distinct bonding characteristics, which is further supported by PDOS results of Si *3p* and O *2p* orbital. On Ag, the PDOS shows limited energetic overlap between Si and O near the Fermi level, pointing to weak orbital hybridization and a less stabilized Si–O bond. In contrast, Si and O states show significantly greater overlap in the -3 to 0 eV region, indicating a stronger electron donation on Pt, Pd, and Ir stabilizes the Si–O linkage and shifts the reaction pathway toward successive Si–C bond breaking. We next examined the selectivity within the Pt, Pd, Au, and Ir group during subsequent breakdown of the *Si–O–Si–$CH_3$ intermediate (purple frame in **Figure 4**). Interestingly, On Pt(111), the Si atom bearing the terminal $CH_3$ group exhibits a highly negative Bader charge (-2.42 e), whereas the corresponding values on Pd(111), Au(111), and Ir(111) are less negative (- 1.97 to -2.19 e). The enhanced charge accumulation on Pt strengthens the Si–O bond *via* increased polarization. Correspondingly, the PDOS shows significantly stronger overlap between Si *3p* and O *2p* orbitals on Pt(111) compared to Pd(111), Au(111), and Ir(111), where the Si–O orbital coupling is noticeably weaker.

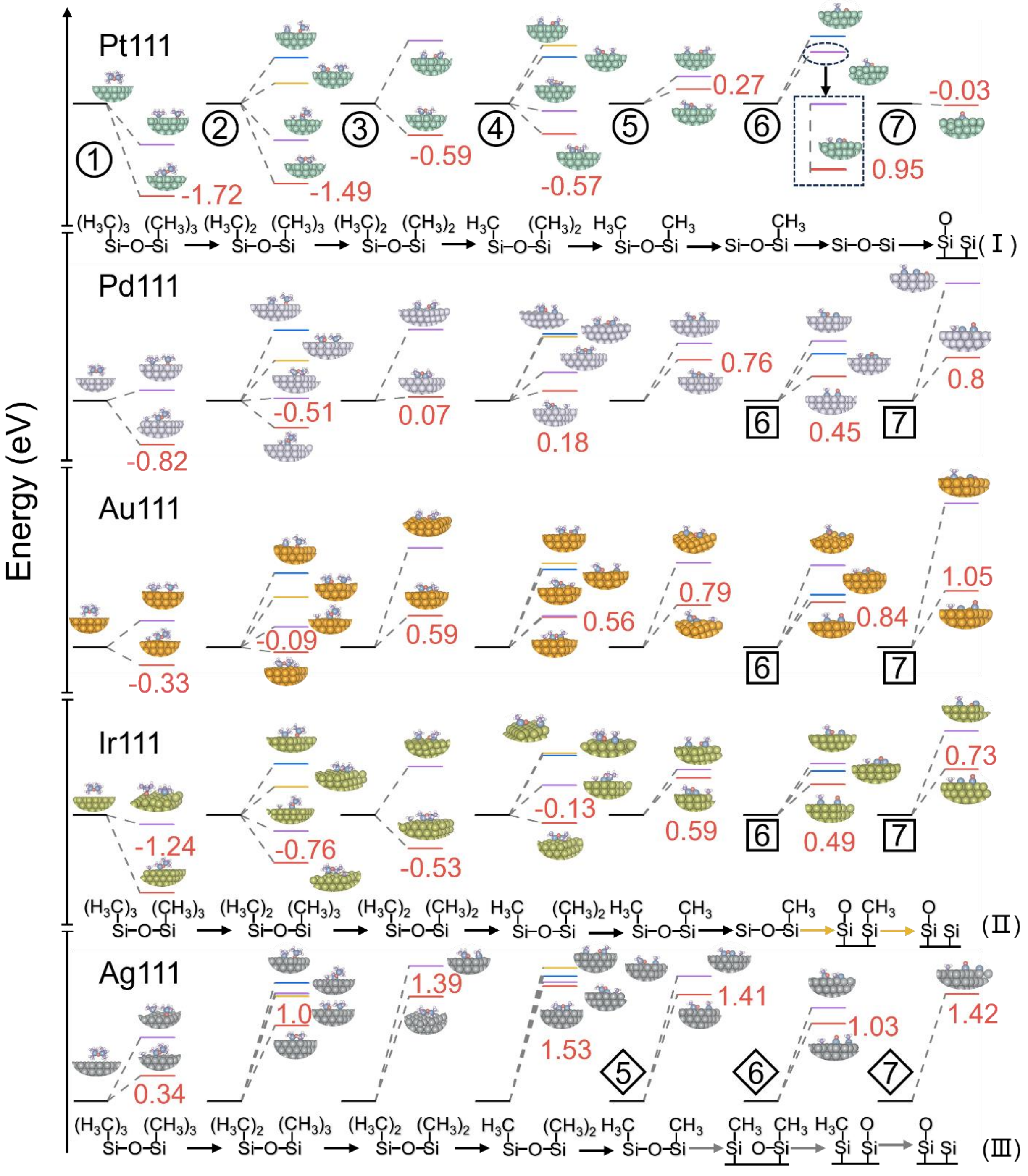

**Figure 3. Decomposition pathways of HMDS.** DFT-calculated decomposition pathways of HMDS on Pt(111), Pd(111), Au(111), Ir(111), and Ag(111), considering every Si-C and Si-O bond breaking possibility, forming surface-bound intermediates such as $^*CH_3$ and *OSi($CH_3$)$_x$. Each step is annotated with its reaction energy (ΔE), defined as the electronic energy difference between the final and initial states. Red stairs denote the most favorable decomposition pathway for each system. In general, three pathways are summarized, which is I for Pt(111), II for Pd(111), Au(111), and Ir(111), and III for Ag(111). For each pathway, seven steps can be identified with the number surrounded by different frames showing the reaction sequence. Color scheme: Si (blue), H (white), C (purple), O (red), Pt (green), Pd (light purple), Au (gold), Ir (olive), and Ag (dark grey).

To determine the kinetic feasibility and rete-determining step (RDS) of the proposed decomposition pathways, we performed climbing-image nudged elastic band (CI-NEB)[20] calculations to evaluate the activation energies associated with key bond-breaking events

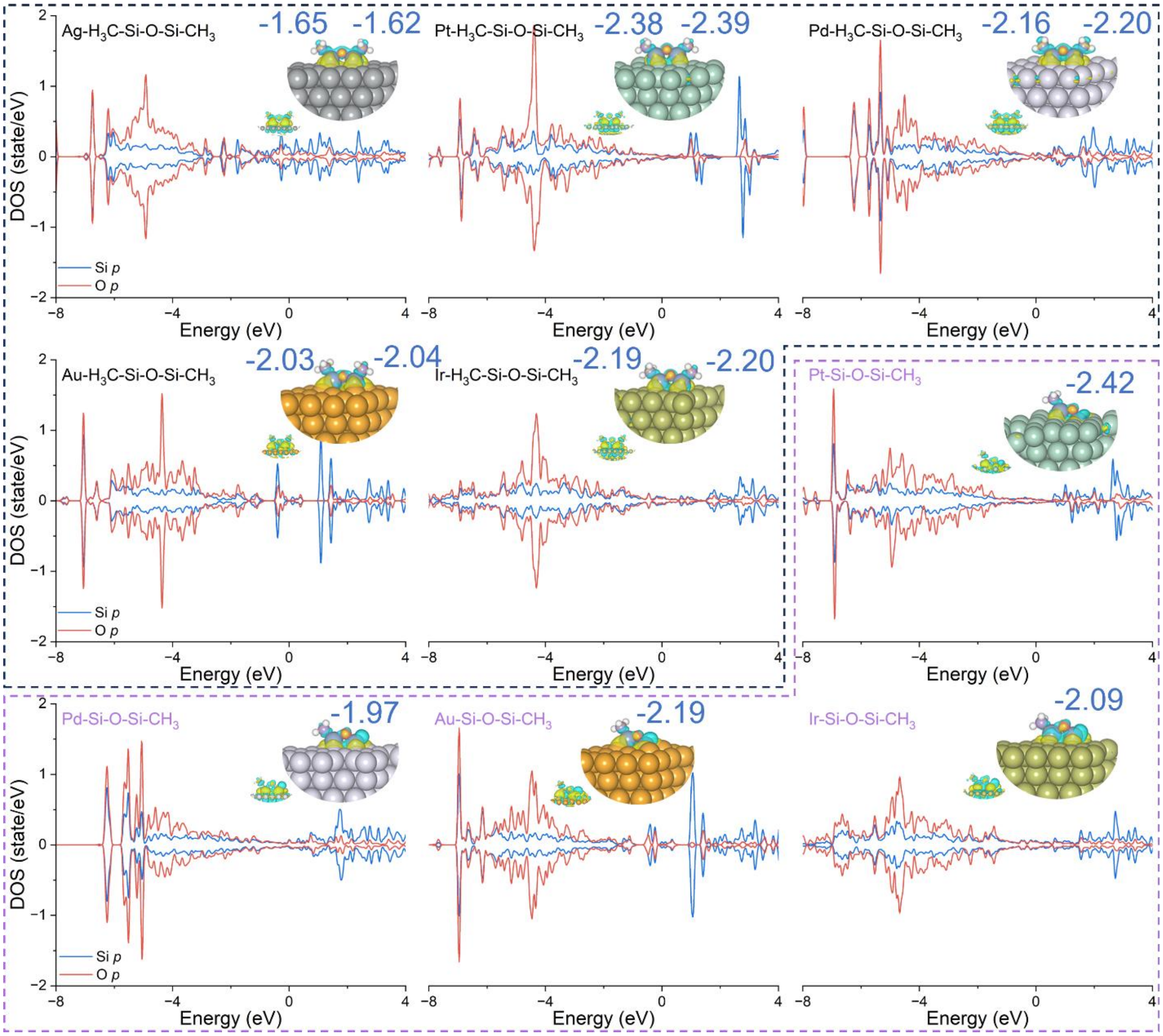

**Figure 4. Electronic structure analyses.** PDOS for the 5th steps (in black frame) and the 6th steps (in purple frame), consistent with the comprehensive decomposition pathway in **Figure 3**. Fermi level was shifted to zero. Insets: Badar charge and charge density difference results on each structure with the number showing the Bader charge value.

Color scheme: Si (blue), H (white), C (purple), O (red), Pt (green), Pd (light purple), Au (gold), Ir (olive), and Ag (dark grey).

on metal surfaces. The energy profiles are resolved into three elementary steps: the first Si–C cleavage (**Figure 5c**), the first Si–O cleavage (**Figure 5d**), and the second Si–C cleavage (**Figure 5e**). Among the thermodynamically preferred Si–C cleavage pathway, the first Si–C bond breaking step exhibits the highest activation barrier, establishing it as the RDS. The second Si–C bond breaking occurs with significantly lower energy barriers, indicating that the molecular framework becomes progressively more reactive following the initial bond rupture. To note, it is rationally assumed here that further Si–C dissociation steps will proceed with even lower or negligible barriers. Therefore, only the first two Si–C cleavages were explicitly treated *via* CI-NEB calculations. Although the Si–O cleavage path shows even higher transition-state energies in some cases, it is thermodynamically disfavored and not part of the dominant reaction route and thus excluded from kinetic relevance. It is worth noting that Ag(111) presents the highest barrier for the first Si–C cleavage of 2.24 eV, suggesting poor reactivity toward HMDS decomposition and a strong resistance to surface poisoning. In contrast, Pt exhibits the lowest barrier of 0.53 eV, indicating high reactivity but also a greater susceptibility to catalyst deactivation *via* siloxane poisoning. Subsequently, the linear scaling relationship between the computed activation energy for the RDS (first Si–C cleavage) and the HMDS adsorption energies on various metal surfaces are examined. As shown in **Figure 5f**, a linear correlation emerges between the two descriptors, with a correlation coefficient ($R^2$) of 0.81. This positive scaling relationship indicates that metals which bind HMDS more strongly tend to lower the energy barrier for Si–C bond activation. Such a trend is consistent with classical Brønsted–Evans–Polanyi (BEP) behavior,[21] where a stronger precursor adsorption leads to stabilization of the transition state.

Given that $^*CH_3$ is a dominant intermediate species generated during HMDS decomposition across multiple metal surfaces, we further investigated its possible fate in the presence of surface oxygen. While previous steps assumed that $^*CH_3$ can be removed under typical sensor operating conditions, we sought to validate this hypothesis by evaluating the thermodynamic feasibility of oxygen promoted $^*CH_3$ dehydrogenation. As shown in **Figure 5e**, $^*CH_3$ adsorption is exothermic across all examined metals, with Pt(111) showing the strongest stabilization. While this suggests surface retention is possible, it does not rule out further reaction, particularly in oxidative environments. To explore this, we modeled *O-assisted dehydrogenation pathways[22] of $^*CH_3$, which closely resemble the initial steps in methane oxidation chemistry,[23, 24] where surface-bound $^*CH_3$ intermediates undergo sequential C–H activation. As illustrated in **Figure 5f**, the $^*CH_3$ group can dehydrogenate to $^*CH_2OH$, *CHOH, and eventually to *C. These steps are thermodynamically favorable on most surfaces, particularly Ag(111), Ir(111), and Au(111), with energy drops up to -0.9 eV. Pt(111), however, presents a slightly endothermic step (+0.18 eV) for the final dehydrogenation. To note, in the realistic gas sensor system, where the operating temperature ranges from room temperature to elevated temperatures (*i.e.,* even up to 400 ºC), the carbon species can undergo further oxidation to $CO_2$, under continued exposure to oxidative conditions. Therefore, the progression from $*CH_3$ to $CO_2$ *via* dehydrogenation and oxidation steps supports the notion that surface-bound methyl fragments do not persist but rather evolve into volatile products that desorb from the surface.

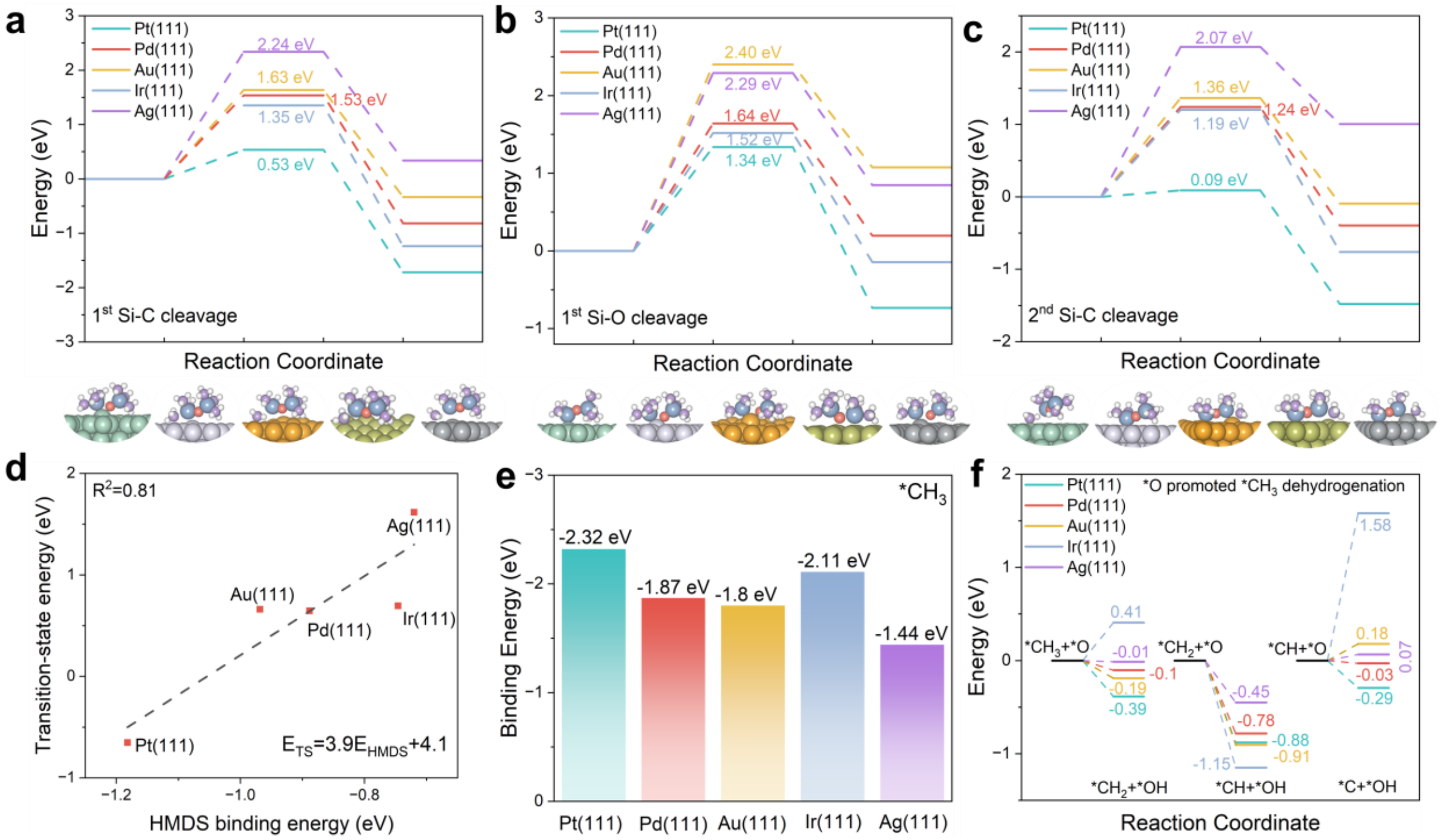

**Figure 5. Kinetic barriers for bond-breaking events during HMDS decomposition and oxygen-promoted $*CH_3$ dehydrogenation pathway. a-c** Activation energies for (a) the first Si–C bond cleavage, (b) the first Si–O bond cleavage second, and (c) the second Si–C bond cleavage. Initial, transition, and final states are denoted by IS, TS, and FS, respectively. Insets below each figure present the atomic structures of every transition state on various metal surfaces. **d,** Linear scaling relation between the transition-state energies and HMDS binding energies for the first Si–C bond cleavage and HMDS adsorption energy on metal surfaces. **e,** Binding energy between metal surfaces and the $*CH_3$ group. **f,** Energy diagrams for O*-promoted $*CH_3$ dehydrogenation. Color scheme: Si (blue), H (white), C (purple), O (red), Pt (green), Pd (light purple), Au (gold), Ir (olive), and Ag (dark grey).

To develop a predictive framework for catalytic activity in HMDS decomposition, we constructed a microkinetic model grounded in surface binding energetics and DFT-calculated transition-state energies. To the best of our knowledge, this is the first-ever development of catalytic volcano model to aid-in the understanding of sensor performance. Considering the working conditions of the catalytic sensor with abundant oxygen, the overall reaction rate is primarily governed by two kinetically relevant processes: Si–C bond cleavage and $O_2$ activation. Each step was mapped to its corresponding adsorption energy, that is HMDS binding energy ($E_{HMDS}$) and oxygen adsorption energy ($E_O$) *via* established scaling relationships. To note, the scaling relations between $O_2$ activation energy and O binding energy on metal surfaces are reproduced based on our previous work within same functional level,[25] as indicated in **Figure S15**. **Figure 6a** provides a comparative view of the adsorption energetics across all five metal catalysts. Binding energies of HMDS, atomic O, transition states, and $^*CH_3$ are displayed in a radar format, illustrating the unique adsorption profiles that distinguish the reactivity of each surface. Using $E_{HMDS}$ and $E_O$ as dual descriptors, we derived a volcano activity map (**Figure 6b**) based upon the *Langmuir* adsorption framework.[26] The reaction rate was computed under steady-state assumptions, accounting for the competitive adsorption of reactants and the rate-limiting nature of bond cleavage and $O_2$ dissociation. Methods relevant to microkinetic modeling are exemplified in **Supplementary Information. Interestingly, the resulting volcano surface captures the trade-off between adsorption strength and reaction turnover: overly weak binding limits activation, while excessively strong binding leads to surface poisoning or site blocking.** Beyond its role in mapping intrinsic catalytic activity, the derived volcano map offers valuable insights for the rational design of siloxane-tolerant sensor materials. Unlike traditional catalyst screening, where the volcano apex is often considered the most desirable, in the context of catalytic sensors operating under long-term exposure to volatile siloxanes, the interpretation is reversed. Materials positioned near the volcano peak, such as Pt(111) and Pd(111), exhibit high activity for HMDS decomposition but are simultaneously more prone to rapid surface poisoning. Their strong adsorption and facile activation of siloxanes can lead to accumulation of decomposition intermediates or silicon-containing residues, resulting in irreversible deactivation. Therefore, in catalytic sensor applications, a balance must be struck between sufficient activity and long-term surface stability. To this end, several synthetic strategies can be envisioned to modulate the surface properties of metal catalysts, like alloying (*e.g.*, incorporating small amounts of Au) or boundary engineering (*e.g.*, defect-rich regions can disrupt extended poisoning pathways by limiting the formation of continuous surface silicate networks). **This interesting volcano framework not only enables activity prediction but also provides practical guidelines for designing siloxane-tolerant sensor materials, highlighting the trade-off between reactivity and surface stability.**

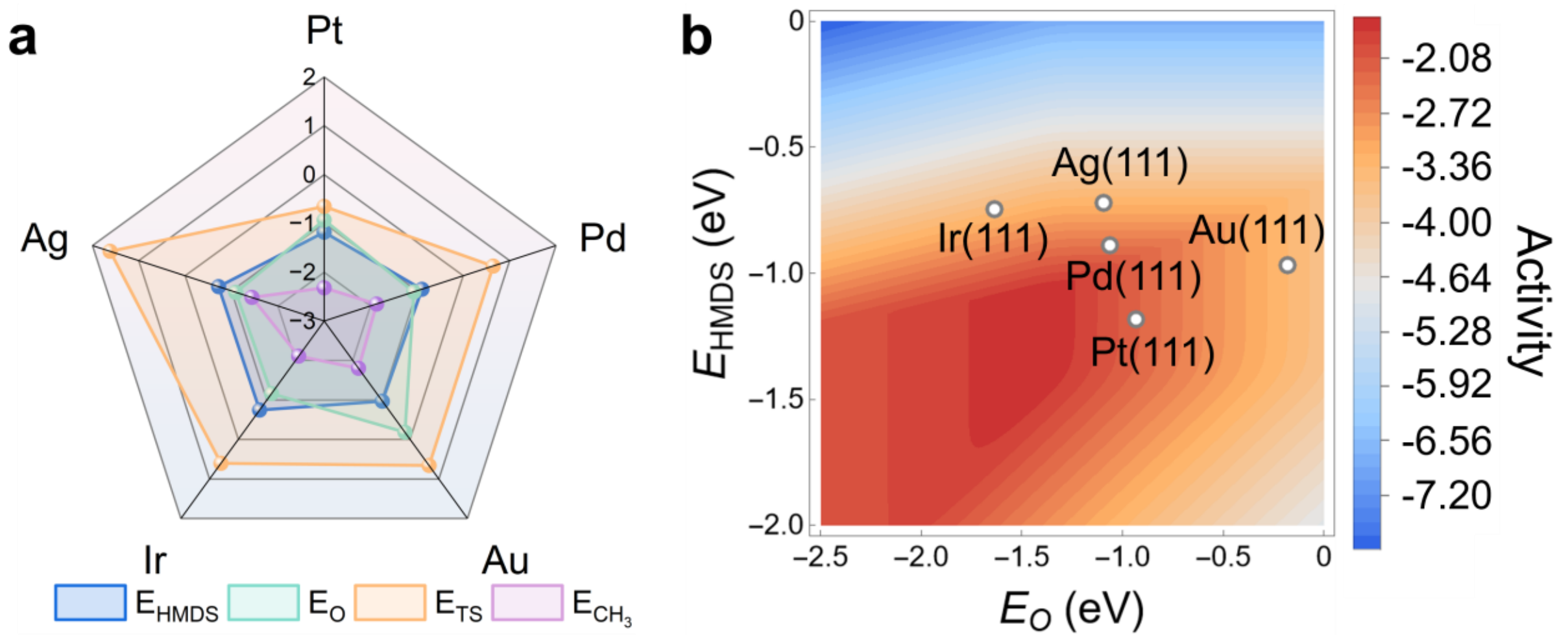

**Figure 6. Microkinetic volcano model for HMDS decomposition on transition metal surfaces. a,** Radar diagram showing the binding energy for different adsorbates involved in this study on various metal systems. All units in eV. **b,** Microkinetic volcano model derived based upon the dual-site Langmuir adsorption framework, using HMDS and O binding energies as the descriptors. The rate is governed by the RDS of Si–C cleavage and $O_2$ activation, each scaled to their respective adsorption energies. DFT-calculated points for five metals are overlaid. The color bar shows the value of $A = kTLog[\frac{r_S h}{KT}]$, as depicted in **Supplementary Information**.

To further validate our computational results, we exposed 440 ppm HMDS vapor to three selected noble metal foils, Pt, Pd, and Au. Prior to HMDS exposure, all foils were electrochemically cleaned to remove organic residues and surface oxides using reported methods.[27] Immediately after cleaning, each foil was flushed with $N_2$ and then exposed to 440 ppm HMDS vapor for 1 hour. The silicon-containing species deposited on the foils were analyzed by XPS (**Figures 7** and **S16**). The high-resolution Si 2*p* spectra after 1 hour of HMDS exposure reveal pronounced $SiO_2$ peaks at ~103.6 eV on Pt and Pd, with the intensity being substantially higher on Pt. In contrast, Au shows only a peak at 102.1 eV attributable to silane species. The estimated surface ratios of deposited silicon species to metal, which were determined from XPS peak areas and relative sensitivity factors (**Table S2**), were 0.217, 0.141, and 0.097 for Pt, Pd, and Au, respectively. These results indicate that Pt exhibits the highest activity toward catalytic HMDS decomposition, likely achieving complete oxidation to $SiO_2$ and leading to substantial silicon deposition. In contrast, HMDS oxidation on Pd and Au appears to be incomplete, terminating primarily at the silane stage. The observed catalytic behavior among Pt, Pd, and Au aligns well with the computational predictions, showing that Si–C bond cleavage proceeds with significantly lower energy barriers on Pt (0.53 and 0.09 eV for the 1$^{st}$ and 2$^{nd}$ Si–C bond cleavage steps, **Figures 5a** and **c**) than on Pd and Au (~1.2–1.6 eV for both steps).

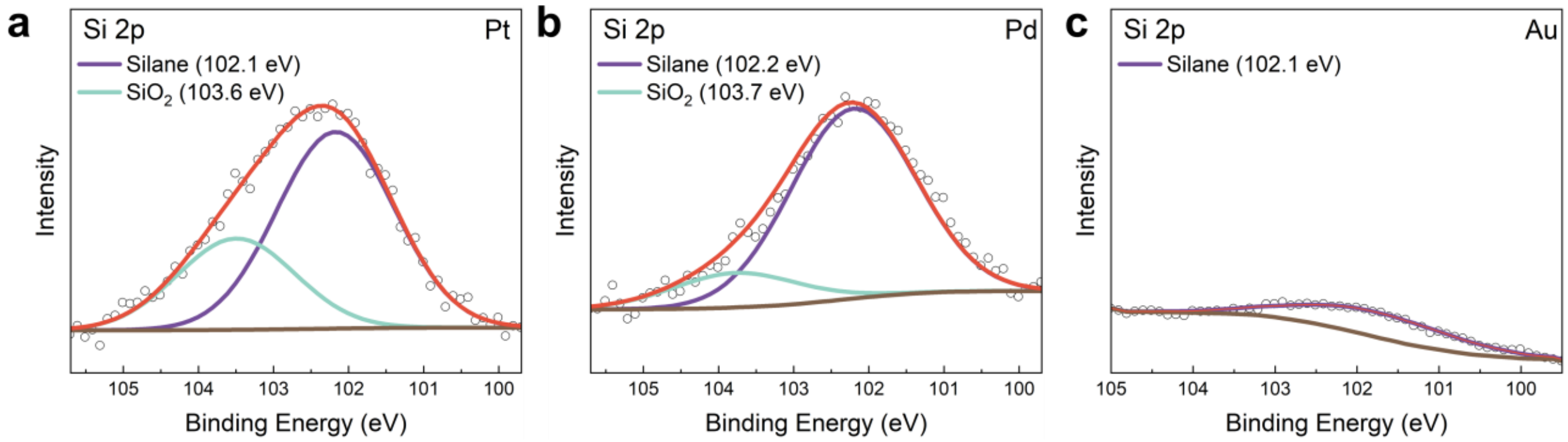

**Figure 7. High-resolution Si 2*p* XPS spectra of a, Pt, b, Pd, and c, Au foils after exposure to 440 ppm HMDS for 1 hour**, showing silicon species deposition on each noble metal and validating the computational results.

To note, these integrated computational and experimental findings not only validate the possible mechanism suggested by AI Agent but also enrich its knowledge base for future iterations. Incorporating these newly obtained data and mechanistic insights back into the AI Agent enables *DigSen* to progressively refine its reasoning, strengthen material–response correlations, and improve the accuracy of predictions for siloxane-related sensor poisoning. This feedback loop between theory, experiment, and AI suggests a promising direction toward self-improving digital discovery in catalytic sensing.

**Taken together, these results also highlight a broader implication: catalytic principles can be directly leveraged to study and understand catalytic sensor degradation.** Building upon the above AI-enhanced insights, the integration of AI-driven discovery, first-principles-based catalysis theory, and experimental validation offers a coherent and generalizable paradigm for diagnosing and mitigating poisoning phenomena across diverse sensing materials. This unified framework provides actionable guidance for the development of next-generation, siloxane-tolerant catalytic sensors and can be readily extended to other material platforms.

## Conclusion

This work establishes a previously unexplored mechanistic understanding of siloxane decomposition on catalytic surfaces and its role in sensor poisoning, guided by a self-developed AI Agent, *DigSen*. By integrating comprehensive theoretical and experimental investigations, we elucidate how HMDS undergoes surface-triggered decomposition to form silicon-containing residues that progressively deteriorate sensing performance. The derived descriptor-based microkinetic volcano model further provides a predictive framework for assessing the susceptibility to siloxane poisoning, offering quantitative guidance for identifying more resilient sensing materials. Integrating these mechanistic insights and experimental data back into *DigSen* enriches its knowledge base and strengthens its ability to reason about siloxane-related degradation, leading to a closed-loop framework**. More broadly, for the first time, this study shows that principles in catalysis can be directly transferred to interpret and predict sensor degradation, and that AI-guided reasoning can accelerate the identification of overlooked materials challenges.** The resulting “AI + theory + experiment” framework thus provides a generalizable paradigm for materials digital discovery, extending far beyond siloxane systems to a wide range of problems from catalyst deactivation to functional materials design.

## Acknowledgement

This work was supported by the JSPS KAKENHI (Nos. JP25K01737 and JP24K23069), and the U.S. Department of Energy (DOE), Office of Science, Office of Basic Energy Sciences (BES): Material Sciences and Engineering Division, Synthesis and Processing Science Program, FWP 78705. We acknowledge the MASAMUNE-IMR at the Center for Computational Materials Science (Institute for Materials Research, Tohoku University) (Nos. 202412-SCKXX-0204 and 202412-SCKXX-0211), and the Institute for Solid State Physics (ISSP) at the University of Tokyo for providing supercomputer resources. L.L.

also gratefully acknowledges support from the University of Utah and the Alfred P. Sloan Foundation (Grant # FH-2023-20829) for supporting the lab setup.

**Author contributions**

H. L. and L. L. mentored, edited, and financed the work. H. L. conducted all the computations, analyzed the data and drafted the manuscript. B. Y did the relevant experiments and participated in the manuscript writing. Y. L. polished the codes of the AI Agent. Y. W. and X. J. helped the microkinetic modeling. All authors discussed the results and contributed to the final manuscript.

**Competing Interest**

The authors declare no Competing interest